\documentclass[aps,prl,twocolumn,groupedaddress]{revtex4-1}
\usepackage{graphicx}
\begin{document}

%\preprint{}

\title{Diagnostics of plasma photoemission at strong coupling}

% Please use the appropriate macro for each type of information.

\author{Babiker Hassanain}
\email[]{babiker@thphys.ox.ac.uk}
%\thanks{}
\affiliation{The Rudolf Peierls Centre for Theoretical Physics,
Department of Physics, University of Oxford. 1 Keble Road,
Oxford, OX1 3NP, UK.}
\author{Martin Schvellinger}
\email[]{martin@fisica.unlp.edu.ar}
%\thanks{}
\affiliation{IFLP-CCT-La Plata, CONICET and
Departamento  de F\'{\i}sica,  Universidad Nacional de La Plata.
 Calle 49 y 115, C.C. 67, (1900) La Plata,  Buenos Aires,
Argentina.}


\date{\today}

\begin{abstract}

We compute the spectrum of photons emitted by the finite-temperature
large-$N$ $SU(N)$ ${\cal {N}}=4$ supersymmetric Yang-Mills plasma
coupled to electromagnetism, at strong yet finite 't Hooft coupling.
We work in the holographic dual description, extended by the
inclusion of the full set of ${\cal{O}}(\alpha'^3)$ type IIB string
theory operators that correct the minimal supergravity action. We find that,
as the t' Hooft coupling decreases, the peak of the spectrum
increases, and the momentum of maximal emission shifts towards the
infra-red, as expected from weak-coupling computations. The total
number of emitted photons also increases as the 't Hooft coupling
weakens.

\end{abstract}


\maketitle

%\section{Introduction}
%\subsection{}
%\subsubsection{}
{\it Introduction.} The analysis of data from heavy ion collision
experiments at RHIC and LHC indicates that the quark-gluon plasmas
(QGPs) produced are in the strongly-coupled regime
\cite{Shuryak:2003xe}, where the 't Hooft coupling governing the
interactions of the microscopic constituents of the plasma is larger
than one. Being electrically-charged, these microscopic constituents
will emit photons. The number of photons emitted with a given
momentum, {\it i.e.} the photoemission spectrum, yields valuable
information about the structure of the plasma. A theoretical study
of this spectrum at strong coupling is therefore an essential step
for investigating the QGP. Gauge/string duality lends itself
perfectly to such a computation, because it allows the investigation
of strongly-coupled gauge theories in terms of their weakly-coupled
supergravity dual description \cite{Maldacena:1997re}. Although
there is no complete string theory dual model which accounts for all
the relevant properties of QCD, the microscopic theory governing the
behavior of plasma produced at RHIC and LHC, one can attempt to
approach the real world using the holographic dual of the large$-N$
$SU(N)$ ${\cal {N}}=4$ super Yang-Mills (SYM) plasma. Moreover,
holography stipulates that $\lambda$ in the gauge theory maps to
$\alpha'^{-2}$ in the string dual, so that the minimal supergravity
({\it i.e.} the zeroth-order string theory description in a
small-curvature expansion), obtained for $\alpha' \to 0$,
corresponds to the gauge theory at $\lambda \rightarrow\infty$.
%{\emph {infinite}} 't Hooft coupling.
With these two caveats in mind, the work of \cite{CaronHuot:2006te}
considered two limits: photoemission from infinitely
strongly-coupled SYM plasma, tackled via minimal type IIB
supergravity on the AdS-Schwarzschild black hole $AdS_{BH} \times
S^5$, and photoemission from weakly-coupled SYM plasma, computed
using the traditional tools of perturbative quantum field theory.
The real-world QGP lies somewhere in between these two illuminating
yet unrealistic regimes. Our aim in this letter is to compute the
photoemission rate of ${\cal {N}}=4$ SYM plasma at large finite
$\lambda$. We work in the holographic dual extended by the inclusion
of the full ${\cal {O}}(\alpha'^3)$ type IIB string theory
corrections to the supergravity action. The relation $\lambda \sim
\alpha'^{-2}$ immediately dictates that the corrections to the
$\lambda \rightarrow \infty$ result start at ${\cal
{O}}(\lambda^{-3/2})$. We compute characteristic
properties of the photoemission spectrum at large finite $\lambda$,
such as the evolution of the height and position of the
photoemission peak as a function of $\lambda$. We thereby quantify
the interpolation between the photoemission spectrum from
strongly-coupled plasma and that from weakly-coupled plasma.

{\it Photoemission rate and spectral function.} ${\cal {N}}=4$ SYM
theory is a supersymmetric gauge theory with gluons, fermions and
scalars all in the adjoint representation of $SU(N)$, and a (global)
$R$-symmetry group $SU(4)$. To model electromagnetism in this
theory, one of the $U(1)$ subgroups of the $R$-symmetry group is
gauged with coupling $e$. The Lagrangian can then be written as
\cite{CaronHuot:2006te}:
\begin{equation}\label{theory}
L = L_{SYM}+e J_\mu^{\small\textrm{em}} {\cal{A}}^\mu
-\frac{1}{4}{\cal{F}}^2 \, ,
\end{equation}
where $L_{SYM}$ is the Lagrangian of ${\cal {N}}=4$ SYM theory, and
the interactions internal to this Lagrangian are governed by the 't
Hooft coupling $\lambda = g_{YM}^2N$. The field ${\cal{A}}^\mu$ is the photon (introduced by hand), with
kinetic term ${\cal{F}}^2$, and $J_\mu^{\small\textrm{em}}$ is the electromagnetic current.
The number of photons produced by a
thermally equilibrated plasma per unit time per unit volume is given
by
\begin{equation}\label{master}
\frac{d\Gamma_\gamma}{dk} = \frac{\alpha_{\small \textrm{em}}}{\pi}
\, k \, n_b(k) \, \eta^{\mu\nu} \, \chi_{\mu\nu}(k) \, ,
\end{equation}
where $k$ is the 3-momentum of the (on-shell) photon,
$n_b(k)=1/(e^{k/T}-1)$, $T$ is the temperature,  and
$\alpha_{\small\textrm{em}}\equiv e^2/4 \pi$. This holds to all
orders in $\lambda$, and to leading order in $e$. The quantity
$\chi_{\mu\nu}(k)$ is the light-like spectral density of the plasma
defined via the retarded electromagnetic current correlator
$R_{\mu\nu}$ as $\chi_{\mu\nu} = -2 \textrm{Im} R_{\mu\nu}$, where
\begin{equation}
R_{\mu\nu}(k) = -i \int d^4x \, e^{i K.X} \, \Theta(t) \,
<[J_\mu^{\small\textrm{em}}(x), J_\nu^{\small\textrm{em}}(0)]> \, ,
\label{Rmunu}
\end{equation}
$\Theta(t)$ is the step function and $K=(k,\vec{k})$, with
$k=|\vec{k}|$.

We wish to compute $R_{\mu\nu}$ using holography, but we do not have
the holographic dual of the theory defined by Eq.(\ref{theory}); we do, however, have the
holographic dual of pure ${\cal {N}}=4$ SYM. The crucial point is
that, to leading order in $e$, the electromagnetic current
$J_\nu^{\small\textrm{em}}=J_\nu$, where $J_\nu$ is purely the
$R$-symmetry current associated with the $U(1)$ subgroup. Therefore,
the two-point function of the electromagnetic current can be
replaced by the two-point function of the $R$-symmetry current,
computed entirely within the ${\cal {N}}=4$ SYM itself
\cite{CaronHuot:2006te}. Our aim in this paper is to compute the
retarded correlator of the $R$-symmetry currents of ${\cal {N}}=4$
SYM at strong 't Hooft coupling, retrieve the spectral function
$\chi_{\mu\nu}$, and insert the latter into Eq.(\ref{master}) to
obtain the photoemission rate. Given that we wish to keep $\lambda$
strong yet finite, we must therefore work in the supergravity dual
to ${\cal {N}}=4$ SYM plasma extended by the addition of finite
$\alpha'$ corrections.

{\it Type IIB string theory setup at ${\cal {O}}(\alpha'^3)$.} The
type IIB supergravity action corrected to ${\cal{O}}(\alpha'^3)$ is
$S_{IIB}=S_{IIB}^{0} + S_{IIB}^{\alpha'}$, where
\begin{equation}
S_{IIB}^{0}=\frac{1}{2 \kappa_{10}^2}\int \, d^{10}x\,
\sqrt{-G}\left[R_{10}-\frac{1}{2}\left(\partial\phi
\right)^2-\frac{1}{4.5!}\left(F_5\right)^2 \right]
\label{action-10D}
\end{equation}
where $\phi$ is the dilaton, $F_5$ the five-form, and $R_{10}$ is
the curvature. The leading 't Hooft coupling corrections are
contained in $ S_{IIB}^{\alpha'}$, and given schematically by
\cite{Paulos:2008tn,Myers:2008yi}
\begin{eqnarray}
\label{10DWeyl}
S_{IIB}^{\alpha'}&=&\frac{\gamma R^6}{2 \kappa_{10}^2}\int \,
d^{10}x\, \sqrt{-G} \, e^{-\frac{3}{2}\phi}
\left(C+{\mathcal{T}}\right)^4 \,, \label{10d-corrected-action}
\end{eqnarray}
where  $\gamma=\frac{1}{8}  \zeta(3) (\alpha'/R^2)^3=\frac{1}{8}
\zeta(3) \lambda^{-3/2}$, with $R^4=g_{YM}^2N\alpha'^2$ and $\zeta$
is the Riemann Zeta function. $C$ is the 10D Weyl tensor, and ${\cal
{T}}_{abcdef}$ is defined as
\begin{eqnarray}
i\nabla_a F^{+}_{bcdef} +\frac{1}{16}\left(F^{+}_{abcmn}F^{+}_{def}{}^{mn}-3
F^{+}_{abfmn}F^{+}_{dec}{}^{mn}\right)\,,\nonumber \label{T-tensor}
\end{eqnarray}
where the RHS is antisymmetrized in $[a,b,c]$ and $[d,e,f]$ and
symmetrized with respect to interchange of $abc \leftrightarrow def$
\cite{Paulos:2008tn}. Also, $F^{+} = \frac{1}{2} (1+\ast) F_5$. The
operators in Eq.(\ref{10d-corrected-action}) are dimension-eight
operators obtained by various independent contractions of $C$ and
${\cal{T}}$ that can be found in \cite{Paulos:2008tn}. The
background solution to this action is the corrected metric
$G_{MN}$\cite{Gubser:1998nz}, with $f(u)=1-u^2$,
\begin{eqnarray}
ds^2 &=& \left(\frac{r_0}{R}\right)^2\frac{1}{u} \, \left(-f(u) \,
K^2(u) \, dt^2 + d\vec{x}^2\right) \nonumber \\
&& + \frac{R^2}{4 u^2
f(u)} \, P^2(u) \, du^2 + R^2 L^2(u) \, d\Omega_5^2 \,
,\label{proper-metric}
\end{eqnarray}
where $d\Omega_5^2$ is the line element on the $S^5$, and
\begin{eqnarray}
K(u) = e^{\gamma \, [a(u) + 4b(u)]}\,, \,\, P(u) = e^{\gamma \,
b(u)}\,, \,\, L(u) = Êe^{\gamma \, c(u)} , \nonumber
\end{eqnarray}
where the exponents are give by the expressions
\begin{eqnarray}
a(u) &=& -\frac{1625}{8} \, u^2 - 175 \, u^4 + \frac{10005}{16} \,
u^6 \, , \nonumber \\
b(u) &=& \frac{325}{8} \, u^2 + \frac{1075}{32} \, u^4
- \frac{4835}{32} \, u^6 \, , \nonumber \\
c(u) &=& \frac{15}{32} \, (1+u^2) \, u^4 \, .
\end{eqnarray}
The extremality parameter is $r_0 = \pi T R^2/(1+265\gamma/16)$,
where $T$ is the physical equilibrium temperature of the plasma. The
boundary of the AdS space is at $u=0$, and the horizon of the black
hole at $u=1$. In addition, both $F_5$ and the dilaton have
non-trivial background solutions, but their explicit forms are of no
consequence in this work, as we explain shortly. The crucial point
is that the tensor ${\cal{T}}$ is zero for the background solution
\cite{Myers:2008yi}.

{\it The vector perturbation.} We now pursue the usual recipe
involved in all holographic computations: firstly, perturb the
supergravity background along the directions $\psi$ which are dual
to the field theory operators ${\cal{J}}$ whose correlation
functions we are interested in, and plug the perturbed background
into $S_{IIB}$. This yields the action ${\cal{S}}(\psi)$ of the
perturbation $\psi$. Then, solve the equations of motion (EOM) of
${\cal{S}}(\psi)$ subject to $\psi = \psi_0$ on the boundary of the
space $u=0$, and evaluate the on-shell action for these solutions,
giving ${\cal{Z}}(\psi_0)$, the generating functional of correlation
functions of the operators ${\cal{J}}$. Differentiating
${\cal{Z}}(\psi_0)$ twice with respect to $\psi_0$ yields
$<{\cal{J}}.{\cal{J}}>$, and we are done. The details of this
prescription in real time are described in \cite{Son:2002sd}. For
the present case the perturbation field $\psi$ dual to the
$R$-symmetry currents $J_\mu$ of the 4D theory is the vector
perturbation $A_\mu$ of the gravitational background obtained as a
solution of the EOM of $S_{IIB}$. The vector
perturbation $A_\mu$ perturbs the metric {\emph{and}} the $F_5$
solution, yielding
\begin{eqnarray}\label{metric-corrected}
&&ds^2=g_{mn} \,dx^m \, dx^n + \nonumber \\
&& R^2 L(u)^2 \, \sum_{i=1}^3 \left[
d\mu_i^2+\mu_i^2(d\phi_i+\frac{2}{\sqrt{3}}A_\mu dx^\mu)^2 \right]
\, ,
\end{eqnarray}
where $g_{mn}=G_{mn}$ for $m,n \in[0,4]$ and $\mu_i$ are the
direction cosines for the sphere, and
\begin{eqnarray}\label{F5-corrected}
F_5 &=& -\frac{4}{R} \overline{\epsilon}+ \frac{R^3
L(u)^3}{\sqrt{3}} \, \left( \sum_{i=1}^3 d\mu_i^2 \wedge d\phi_i
\right) \wedge \overline{\ast} F_2 \,,
\end{eqnarray}
where $F_2 = dA$ is the Abelian field strength of $A_\mu$ and
$\overline{\epsilon}$ is a deformation of the volume form of the
metric of the $AdS_{BH}$. The Hodge duals $\ast$ and
$\overline{\ast}$ are taken with respect to the 10D metric and
5D-$AdS_{BH}$ metric, respectively. Notice that the $\gamma \to 0$
limits of Eq.(\ref{metric-corrected}, \ref{F5-corrected}) are known
exactly, see for instance \cite{Chamblin}. Both of
Eq.(\ref{metric-corrected}, \ref{F5-corrected}) are correct to
linear order in $\gamma$. We will insert these {\it
Ans${\ddot{a}}$tze} into $S_{IIB}=S_{IIB}^{0} + S_{IIB}^{\alpha'}$
below, obtaining an effective Lagrangian for $A_\mu$ which is at
most {\emph{quadratic}} in $A_\mu$.
%(because we are after the two-point functions of $J_\mu$).
Two important points must be stated to this end, dictated by the
fact that we work strictly to linear order in $\gamma$. For
insertion into $S_{IIB}^{0}$, we require the {\it
Ans${\ddot{a}}$tze} to linear order in $\gamma$. However,
$S_{IIB}^{0}$ only contains quadratic powers of $F_5$ and,
therefore, the $\overline{\epsilon}$ part of the $F_5$ Ansatz
{\emph{cannot}} contribute to the quadratic effective Lagrangian of
$A_\mu$. On the other hand, $S_{IIB}^{\alpha'}$ contains operators
which are higher than quadratic in $F_5$, so here
$\overline{\epsilon}$ can contribute to the quadratic Lagrangian for
$A_\mu$, but the crucial point is that $S_{IIB}^{\alpha'}$ contains
an explicit factor of $\gamma$ already, and so we only require
$\overline{\epsilon}$ to zero order in $\gamma$. This is of course
nothing but the volume form on the AdS space \cite{Chamblin}. The
incredibly simplifying upshot of these observations is that
{\emph{we do not require the ${\cal{O}}(\gamma)$ terms in
$\overline{\epsilon}$ for our computation}}. This is why we do not
care to state the explicit form of $\overline{\epsilon}$. All we
need to know is $\textrm{lim}_{\gamma \to 0}\overline{\epsilon}$.
The same observations can be made for the contribution to the
effective Lagrangian of $A_\mu$ of operators containing the dilaton
(and in fact any other field which is trivial in the zero order
background supergravity solution). These statements make the
following work possible.

{\it The effective Lagrangian of the vector perturbation.} Without
loss of generality, we may set the photon momenta to $(k,0,0,k)$. In
order to study the photoemission rate we only need to consider the
transverse fluctuation $A_x(t, x, u)$. Inserting the {\it
Ans${\ddot{a}}$tze} of Eq.(\ref{metric-corrected},
\ref{F5-corrected}) into $S_{IIB}$, and integrating out the $S^5$,
we obtain a complicated Lagrangian for $A_x(t, x, u)$. This action
can be coverted into a more useful form by use of the following
field redefinitions: write $\Psi(u) = A_k(u)/(\sqrt{f(u)}[1+\gamma
p(u)])$, where $p(u)$ is a polynomial in $u$ beginning at
${\cal{O}}(u^2)$ and $A_k(u)$ is the Fourier transform of $A_x(t, x,
u)$. This takes us into the Schr\"odinger basis, such that the
action is given by
\begin{eqnarray}
S &=&- \frac{N^2 r_0^2}{16 \pi^2 R^4} \int \frac{d^4k}{(2 \pi)^4}
\int_0^1 du \, \left[ \frac{1}{2}\Psi {\mathcal{L}} \Psi
+\,\partial_u \Phi \right]  \label{ActionOnShell}
\end{eqnarray}
where ${\mathcal{L}} \, \Psi(u) = 0$ is the EOM
$\Psi''(u)=V(u)\Psi(u)$, where the Schr\"odinger-like potential is
given by
\begin{eqnarray}\label{potential}
V(u) &=& -\frac{1}{f^2(u)}\bigg(1+q^2 u - \frac{\gamma}{144} f(u)
\left[ -11700\right. \nonumber \\
&& \left. +2098482u^2 -4752055 u^4 +1838319 u^6 +q^2 u \right. \nonumber \\
&& \left.  \left. \times (-16470+245442u^2+1011173 u^4
\right)\right]\bigg) \,,
\end{eqnarray}
and $q=k/(2\pi T)$. The boundary term can be simplified to %
%\begin{eqnarray}
%
$\Phi=\Psi'(u)\Psi(u) \, .\label{finalb}$
%
%\end{eqnarray}
We solve the Schr\"odinger equation analytically for the region $k
\ll T$, the so-called hydrodynamic regime of the plasma, and the
high-energy regime $k \gg T$. In the intermediate momentum region we
resort to a numerical solution of the EOM. Once we have the solution
of the Schr\"odinger equation, the trace of the spectral function is
given by:
\begin{equation}\label{EOM-chi}
\chi^{\mu}_{\mu}(k)
=\frac{N^2T^2}{2}\,\left(1-\frac{265}{8}\gamma\right)\textrm{Im}
\left.\left[\frac{\Psi'(u)}{\Psi(u)}\right] \right\vert_{u=0} \, ,
\end{equation}
with $\left(1-\frac{265}{8}\gamma\right)$ coming from the factors of
$r_0$ in Eq.(\ref{ActionOnShell}).

{\it The EOM of the vector perturbation.} We solve the Schr\"odinger
equation using perturbation theory. Write $\Psi(u)=\Psi_0(u)+\gamma
\Psi_1(u)$, and insert into the Schr\"odinger equation, separating
the powers of $\gamma$. The equation for $\Psi_0(u)$ is given by
$\Psi_0''(u)=\left(-f^{-2}(u)(1+q^2 u) \right)\Psi_0(u)$, and solves
to give
\begin{eqnarray}
&& \Psi_0(u) = (1-u)^{-\frac{1}{2}(1+i q)}(1+u)^{-\frac{1}{2}(1+ q)} \nonumber \\
&& {}_2F_1\left(1-\frac{(1+i)q}{2},-\frac{(1+i)q}{2}, 1-iq, \frac{1-u}{2} \right) \,.
\end{eqnarray}
The equation for $\Psi_1(u)$ is solved numerically (if necessary).
The trace of the spectral function $\chi^{\mu}_{\mu}(k)$ is
\begin{eqnarray}\label{chi-corrected}
\frac{N^2T^2}{2}\textrm{Im}\left.\left[\left(1-\frac{265}{8}\gamma\right)
\frac{\Psi'_0}{\Psi_0}+\gamma
\left[-\frac{\Psi'_0}{\Psi_0}\frac{\Psi_1}{\Psi_0}+\frac{\Psi'_1}{\Psi_0}
\right]\right]\right\vert_{u=0} \nonumber
\end{eqnarray}
which is exact to linear order in $\gamma$. We note that
$\chi^{\mu}_{\mu}(k)$ at $\lambda \rightarrow \infty$ is known
\cite{CaronHuot:2006te}, so our task here is to compute the 't Hooft
coupling corrections to that result.

{\it Asymptotics of the spectral function.} $\chi^{\mu}_{\mu}(k)$
% The trace of the spectral function
can be evaluated analytically for low- and high-momentum, and
numerically for the remaining momentum domain. We do not discuss
the details of the computations, referring the reader to
\cite{CaronHuot:2006te}, and we simply display the results:
\begin{equation}\label{chi-limit}
\frac{\chi^{\mu}_{\mu}(q)}{\frac{1}{2}N^2T^2} = \left\{ \begin{array}{cc}
\left(1+\frac{14993}{9}\gamma \right)q +{\cal{O}}(q^3) & \, q \ll 1 \\
\frac{3^{5/6}}{2}
\frac{\Gamma\left(\frac{2}{3}\right)}{\Gamma\left(\frac{1}{3}\right)}\left(1
+5\gamma\right)q^{2/3}+{\cal{O}}(1)& \, q\gg 1 \end{array} \right. \,.
\end{equation}
The coefficient of $q$ in the low-$q$ regime of Eq.(\ref{chi-limit})
means that the electrical conductivity of the strongly-coupled
plasma is enhanced by a factor $\left(1+\frac{14993}{9}\gamma
\right)$ due to the finite $\lambda$ corrections
\cite{Hassanain:2010fv}. This is as expected from the perturbative
computations in \cite{CaronHuot:2006te}: the weakly-coupled plasma
has a larger mean-free-path per collision, allowing more efficient
diffusion of electric charge, and hence a higher electrical
conductivity. The high-$q$ region however, poses a question that we
(as of yet) cannot answer: the authors of \cite{CaronHuot:2006te}
claim that the spectral function at weak coupling should go like
$q^{1/2}$ in the ultraviolet (UV). Given that the spectral function
at $\lambda \rightarrow \infty$ goes like $q^{2/3}$, in that regime
one would have expected our result in Eq.(\ref{chi-limit}) to
display some smooth interpolation between $q^{1/2}$ and $q^{2/3}$.
We do not obtain such an interpolation, finding instead that the
finite coupling corrections {\emph{do not}} change the
$q$-dependence in the UV. Moreover, we find an enhancement by a
factor $(1+5\gamma)$ in that regime (see also
\cite{Hassanain:2009xw}). The fact that the leading $q$ behavior is
unchanged by the corrections could have been seen from the
Schr\"odinger-like potential in Eq.(\ref{potential}): the only
$q$-dependence is $q^2$, identically to the $\lambda \rightarrow
\infty$ case. Terms like $q^4$, which could have changed the
high$-q$ functional dependence of $\chi^\mu_\mu(q)$, vanish. We
shall revisit this point below.
\begin{figure}\label{emission}
\begin{center}
\includegraphics{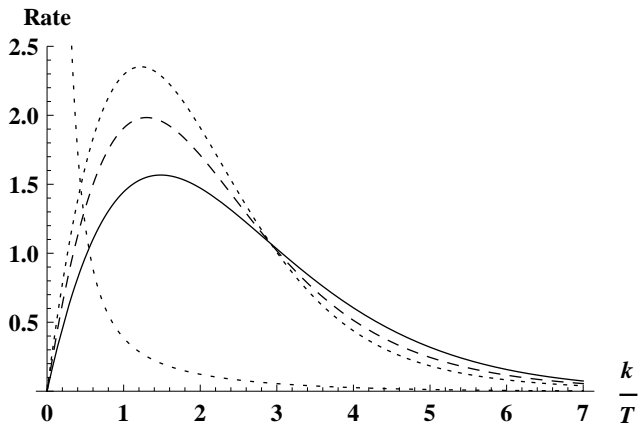}
\caption{The photoemission rate $d\Gamma_\gamma/dk$ in units of
$0.01\alpha_{\small \textrm{em}} N^2 T^3$ for different values of
$\lambda$, as a function of $k/T$.
%, where $k$ is the momentum of the photon.
% and $T$ is the plasma equilibrium temperature.
Solid,
dashed, and small-dashed lines correspond to decreasing values of
$\lambda \rightarrow \infty, 75$ and 50, respectively. The dotted
line to the extreme left is the weak-coupling result at $\lambda =
0.5$ taken from \cite{CaronHuot:2006te}. \label{fig-photoemission}}
\end{center}
\end{figure}

{\it The photoemission spectrum.} We plug the obtained
$\chi^{\mu}_{\mu}(k)$ into Eq.(\ref{master}) to give the
photoemission spectrum. We display the results in figure 1. Clearly,
the corrected result at strong coupling approaches the
weakly-coupled result (taken from reference
\cite{CaronHuot:2006te}). Firstly, the corrected curves exhibit a
steeper tangent at the origin, due to the enhancement of the
electric conductivity by the factor $\left(1+\frac{14993}{9}\gamma
\right)$ in Eq.(\ref{chi-limit}). Secondly, the peak of the
photoemission is enhanced by the corrections, and the momentum of
maximal emission shifts towards the infrared (IR), taking the
corrected curves closer to the weakly coupled result. Simple
numerical analysis on the light-like spectral function yields that
the maximal rate is given by
\begin{equation}
\left. \frac{d\Gamma_\gamma}{dk} \right\vert_{max} \simeq
0.0156695\left( 1 +\left[ 1115.3 - \frac{265}{8}\right]\gamma
\right) + {\cal {O}}(\gamma^2) \, ,
\end{equation}
in units of $\alpha_{\small \textrm{em}} N^2 T^3$, where we have
made explicit the factor $-265/8 \gamma$ coming from the overall
normalization of the action. The position of the peak $k_{max}$ is
\begin{equation}
k_{max} \simeq 1.48469\left( 1 - 842.425 \gamma  \right) T + {\cal
{O}}(\gamma^2) \, .
\end{equation}
This quantity can analytically be shown to be
independent of the overall normalization of the action, making it an
excellent candidate for comparing disparate gauge theories. One more
quantity which is of interest is the total number of photons
emitted, given by the area under the curves in figure 1. This is
enhanced by a factor
\begin{equation}
\frac{N_{\small\textrm{total}}(\gamma)}{N_{\small\textrm{total}}(0)}
\simeq 1 + \left[ 461.941 - \frac{265}{8} \right] \gamma + {\cal
{O}}(\gamma^2) \, ,
\end{equation}
due to the fact that the weakly-coupled theory dominates in the IR,
where Bose-suppression (due to $n_b(k)$) is small.

We finally make two comments about the behavior of the photoemission
rate for high $q$. Firstly, there is a ($\lambda$-independent)
crossover point around $k/T \sim 2.92$, where the corrected curves
dip below the $\lambda \rightarrow \infty$ result. This is expected
from the weak-coupling computations of \cite{CaronHuot:2006te}. What
is surprising, as we mentioned above, is that the asymptotic values
of the $\lambda$-corrected curves for large $k/T$ are given by
$(1+5\gamma)$ times the infinite coupling result, as in Eq.(\ref{chi-limit})
(note that the domain of figure 1 does not extend
to cover this asymptotic behaviour). This means that the
finite-$\lambda$ corrections {\emph{enhance}} the photoemission rate
in the deep UV regime, contrary to the expectations of
\cite{CaronHuot:2006te}. Obviously, we are not guaranteed that the
weakly-coupled result should be approached by strongly-coupled
corrections computed in perturbation theory, especially not for a
situation where the functional dependence on momenta is expected to
be different, so we are not unduly concerned by this apparent
discrepancy. It would be very revealing to understand these
cross-over points, as well as their scaling with $\lambda$. An important extension of our work
would be to determine if the universality found in \cite{Buchel:2008ae} for the energy-momentum spectral functions
operates for the $R$-symmetry current spectral function computed here.

\begin{acknowledgments}
We thank Gert Aarts, Alex Buchel, Carlos N\'u\~nez, Miguel Paulos
and Jorge Russo for valuable comments. The work of M.S. has been
partially supported by the CONICET, the ANPCyT-FONCyT Grant
PICT-2007-00849, and the CONICET Grant PIP-2010-0396.
\end{acknowledgments}

\bibliography{}

\begin{thebibliography}{99}

%\cite{Shuryak:2003xe}
\bibitem{Shuryak:2003xe}
  E.~Shuryak,
%  ``Why does the quark gluon plasma at RHIC behave as a nearly ideal fluid?,''
  Prog.\ Part.\ Nucl.\ Phys.\  {\bf 53} (2004) 273
  [arXiv:hep-ph/0312227].
  %%CITATION = PPNPD,53,273;%%
%\bibitem{Schukraft:2011kc}
  J.~Schukraft and f.~t.~A.~Collaboration,
%  ``First Results from the ALICE experiment at the LHC,''
  arXiv:1103.3474 [hep-ex].
  %%CITATION = ARXIV:1103.3474;%%
%\cite{Abreu:2007kv}
%\bibitem{Abreu:2007kv}
  N.~Armesto {\it et al.},
%  ``Heavy Ion Collisions at the LHC - Last Call for Predictions,''
  J.\ Phys.\ G {\bf 35} (2008) 054001
  [arXiv:0711.0974 [hep-ph]].
  %%CITATION = JPHGB,G35,054001;%%

%\cite{Maldacena:1997re}
\bibitem{Maldacena:1997re}
  J.~M.~Maldacena,
%  ``The large N limit of superconformal field theories and supergravity,''
  Adv.\ Theor.\ Math.\ Phys.\  {\bf 2} (1998) 231
  [Int.\ J.\ Theor.\ Phys.\  {\bf 38} (1999) 1113]
  [arXiv:hep-th/9711200].
  %%CITATION = IJTPB,38,1113;%%

%\cite{CaronHuot:2006te}
\bibitem{CaronHuot:2006te}
  S.~Caron-Huot et al,
%  P.~Kovtun, G.~D.~Moore, A.~Starinets and L.~G.~Yaffe,
%  ``Photon and dilepton production in supersymmetric Yang-Mills plasma,''
  JHEP {\bf 0612} (2006) 015
  [arXiv:hep-th/0607237].
  %%CITATION = JHEPA,0612,015;%%

%\cite{Myers:2008yi}
\bibitem{Myers:2008yi}
  R.~C.~Myers, M.~F.~Paulos and A.~Sinha,
%  ``Quantum corrections to eta/s,''
  Phys.\ Rev.\  D {\bf 79} (2009) 041901
  [arXiv:0806.2156 [hep-th]].
  %%CITATION = PHRVA,D79,041901;%%

%\cite{Paulos:2008tn}
\bibitem{Paulos:2008tn}
  M.~F.~Paulos,
%  ``Higher derivative terms including the Ramond-Ramond five-form,''
  JHEP {\bf 0810} (2008) 047
  [arXiv:0804.0763 [hep-th]].
  %%CITATION = JHEPA,0810,047;%%

%\cite{Gubser:1998nz}
\bibitem{Gubser:1998nz}
  S.~S.~Gubser, I.~R.~Klebanov and A.~A.~Tseytlin,
%  ``Coupling constant dependence in the thermodynamics of N = 4  supersymmetric
%  Yang-Mills theory,''
  Nucl.\ Phys.\  B {\bf 534} (1998) 202.
  %[arXiv:hep-th/9805156].
  %%CITATION = NUPHA,B534,202;%%
%
%\cite{Pawelczyk:1998pb}
%\bibitem{Pawelczyk:1998pb}
  J.~Pawelczyk and S.~Theisen,
%  ``AdS(5) x S(5) black hole metric at O(alpha'**3),''
  JHEP {\bf 9809} (1998) 010.
  %[arXiv:hep-th/9808126].
  %%CITATION = JHEPA,9809,010;%%
  %
  %\cite{Buchel:2004di}
%\bibitem{Buchel:2004di}
  A.~Buchel, J.~T.~Liu and A.~O.~Starinets,
  %``Coupling constant dependence of the shear viscosity in N=4 supersymmetric
  %Yang-Mills theory,''
  Nucl.\ Phys.\  B {\bf 707} (2005) 56
  %[arXiv:hep-th/0406264].
  %%CITATION = NUPHA,B707,56;%%

 %\cite{Son:2002sd}
\bibitem{Son:2002sd}
  D.~T.~Son and A.~O.~Starinets,
  %``Minkowski-space correlators in AdS/CFT correspondence: Recipe and applications,''
  JHEP {\bf 0209} (2002) 042
  [arXiv:hep-th/0205051].
  %%CITATION = JHEPA,0209,042;%%
%\cite{Policastro:2002se}
%\bibitem{Policastro:2002se}
  G.~Policastro, D.~T.~Son and A.~O.~Starinets,
  %``From AdS/CFT correspondence to hydrodynamics,''
  JHEP {\bf 0209} (2002) 043
  [arXiv:hep-th/0205052].
  %%CITATION = JHEPA,0209,043;%%

%\cite{Chamblin:1999tk}
\bibitem{Chamblin}
 A.~Chamblin {\it et al.}
%R.~Emparan, C.~V.~Johnson and R.~C.~Myers,
 %``Charged AdS black holes and catastrophic holography,''
 Phys.\ Rev.\ ÊD {\bf 60} (1999) 064018
 [arXiv:hep-th/9902170].
%%CITATION = PHRVA,D60,064018;%%
%\cite{Cvetic:1999xp}
%\bibitem{Cvetic:1999xp}
 M.~Cvetic, {\it et al.}
%M.~J.~Duff, P.~Hoxha, J.~T.~Liu, H.~Lu, J.~X.~Lu, R.~Martinez-Acosta, C.~N.~Pope {\it et al.},
%``Embedding AdS black holes in ten-dimensions and eleven-dimensions,''
 Nucl.\ Phys.\ Ê{\bf B558}, 96-126 (1999) [hep-th/9903214].

%\cite{Hassanain:2010fv}
\bibitem{Hassanain:2010fv}
  B.~Hassanain and M.~Schvellinger,
%  ``Towards 't Hooft parameter corrections to charge transport in
%  strongly-coupled plasma,''
  JHEP {\bf 1010} (2010) 068
  [arXiv:1006.5480 [hep-th]].
  %%CITATION = JHEPA,1010,068;%%
  %\cite{Hassanain:2011fn}
%\bibitem{Hassanain:2011fn}
  B.~Hassanain and M.~Schvellinger,
  %``Plasma conductivity at finite coupling,''
  arXiv:1108.6306 [hep-th].
  %%CITATION = ARXIV:1108.6306;%%

%\cite{Hassanain:2009xw}
\bibitem{Hassanain:2009xw}
  B.~Hassanain and M.~Schvellinger,
%  ``Holographic current correlators at finite coupling and scattering off a
%  supersymmetric plasma,''
  JHEP {\bf 1004} (2010) 012
  [arXiv:0912.4704 [hep-th]].
  %%CITATION = JHEPA,1004,012;%%

%\cite{Buchel:2008ae}
\bibitem{Buchel:2008ae}
  A.~Buchel, {\it et al.}
  %``Universal holographic hydrodynamics at finite coupling,''
  Phys.\ Lett.\  B {\bf 669} (2008) 364.
  %[arXiv:0808.1837 [hep-th]].
  %%CITATION = PHLTA,B669,364;%%

\end{thebibliography}

\end{document}